\documentclass[letterpaper,twocolumn,10pt]{article}
\usepackage{epsfig,endnotes}
\usepackage{color}
\usepackage{verbatim}
\usepackage{listings}
\usepackage{subfigure}
\usepackage{enumitem}
\usepackage{times}
\usepackage[hidelinks]{hyperref}
\usepackage{amssymb}
\usepackage{pifont}
\usepackage{url}
\usepackage{xcolor}

\newcommand{\secref}[1]{Section~\ref{#1}}
\newcommand{\figref}[1]{Figure~\ref{#1}}

\newcommand{\fixmec}[2]{{\color{#1}{[#2]}}}

\definecolor{royalblue}{RGB}{65,105,225}

\newif\ifcommenton
\commentontrue
\ifcommenton
\newcommand{\alexey}[1]{\fixmec{royalblue}{\bf [AT: #1]}}
\newcommand{\robert}[1]{\fixmec{red}{\bf [RN: #1]}}
\newcommand{\philipp}[1]{\fixmec{green}{\bf [PM: #1]}}
\newcommand{\steph}[1]{\fixmec{brown}{\bf [SW: #1]}}
\newcommand{\jss}[1]{\fixmec{magenta}{\bf [JSS: #1]}}
\newcommand{\ion}[1]{\fixmec{blue}{\bf [IS: #1]}}
\else
\newcommand{\alexey}[1]{}
\newcommand{\robert}[1]{}
\newcommand{\philipp}[1]{}
\newcommand{\steph}[1]{}
\newcommand{\jss}[1]{}
\newcommand{\ion}[1]{}
\fi

\newcommand\blfootnote[1]{%
  \begingroup
  \renewcommand\thefootnote{}\footnote{#1}%
  \addtocounter{footnote}{-1}%
  \endgroup
}

\begin{document}

\date{}

\title{\Large \bf Real-Time Machine Learning: The Missing Pieces}

\author{
{\rm Robert Nishihara$^{*}$\blfootnote{equal contribution}, Philipp Moritz$^{*}$, Stephanie Wang, Alexey Tumanov, William Paul,}
\and
{\rm Johann Schleier-Smith, Richard Liaw, Mehrdad Niknami, Michael I.~Jordan, Ion Stoica}
\\
\textit{UC Berkeley}
} 

\maketitle

\thispagestyle{empty}

\subsection*{Abstract}

Machine learning applications are increasingly deployed not only to serve
predictions using static models, but also as tightly-integrated components of
feedback loops involving dynamic, real-time decision making.
These applications pose a new
set of requirements, none of which are difficult to achieve in isolation, but
the combination of which creates a challenge for existing distributed execution
frameworks: computation with millisecond latency at high throughput,
adaptive construction of arbitrary task graphs, and execution of heterogeneous
kernels over diverse sets of resources. We assert that a new distributed
execution framework is needed for such ML applications and propose a candidate
approach with a proof-of-concept architecture that achieves a 63x performance
improvement over a state-of-the-art execution framework for a representative
application.

\thispagestyle{empty}
\section{Introduction}
\label{sec:intro}

\begin{figure}[t]
    \centering
    \includegraphics[width=0.8\columnwidth]{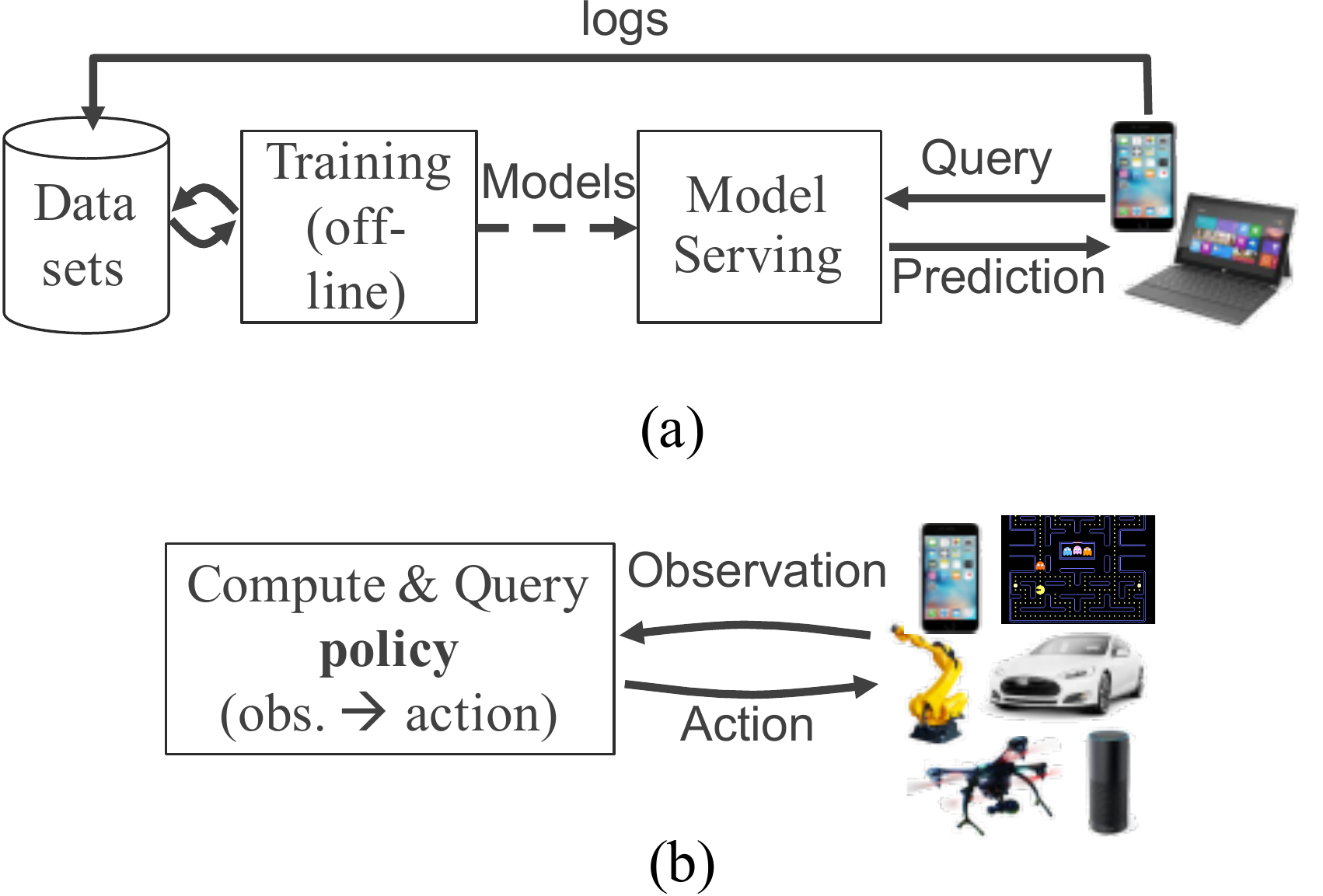}
    \caption{\small{
    	(a) Traditional ML pipeline (off-line training). (b) Example reinforcement learning pipeline: the system continously interacts with an environment to learn a policy, i.e., a mapping between observations and actions.
    }
    }
    \label{fig:rl-models}
\end{figure}

The landscape of machine learning~(ML) applications is undergoing a significant
change. While ML has predominantly focused on training and serving
predictions based on static models (\figref{fig:rl-models}a), there is now a strong shift toward the tight
integration of ML models in feedback loops.
Indeed, ML applications are expanding from the supervised learning paradigm, in
which static models are trained on offline data, to a broader paradigm,
exemplified by reinforcement learning (RL), in which
applications may operate in real environments,
fuse and react to sensory data from numerous input streams, perform continuous
micro-simulations, and close the loop by taking actions that affect
the sensed environment (\figref{fig:rl-models}b).

Since learning by interacting with the real world can be unsafe, impractical, or
bandwidth-limited, many reinforcement learning systems rely
heavily on {\bf simulating physical or virtual environments}. Simulations may be used
during training (e.g., to learn a neural network policy), and during deployment.
In the latter case, we may constantly update the simulated environment as we
interact with the real world and perform many simulations to figure out the
next action (e.g., using online planning algorithms like Monte Carlo tree
search). This requires the ability to perform simulations faster than real
time.

\begin{figure*}[tbh]
\begin{center}
    \subfigure[multiple sensor inputs]{\includegraphics[width=0.3\textwidth]{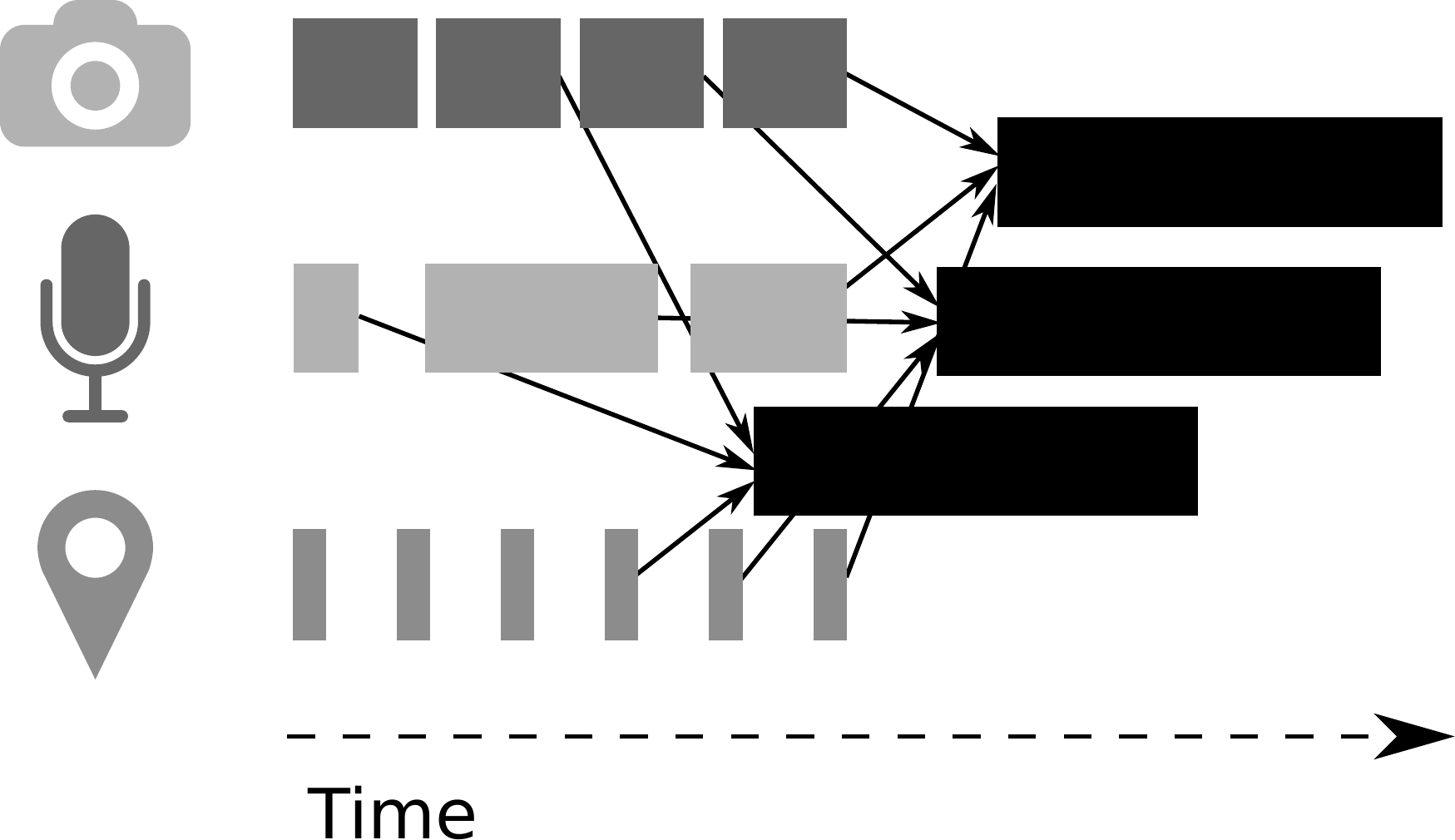}\label{fig:motivation:sensor}}
    \hspace{0.03\textwidth}
    \subfigure[Monte Carlo tree search (MCTS)]{\includegraphics[width=0.3\textwidth]{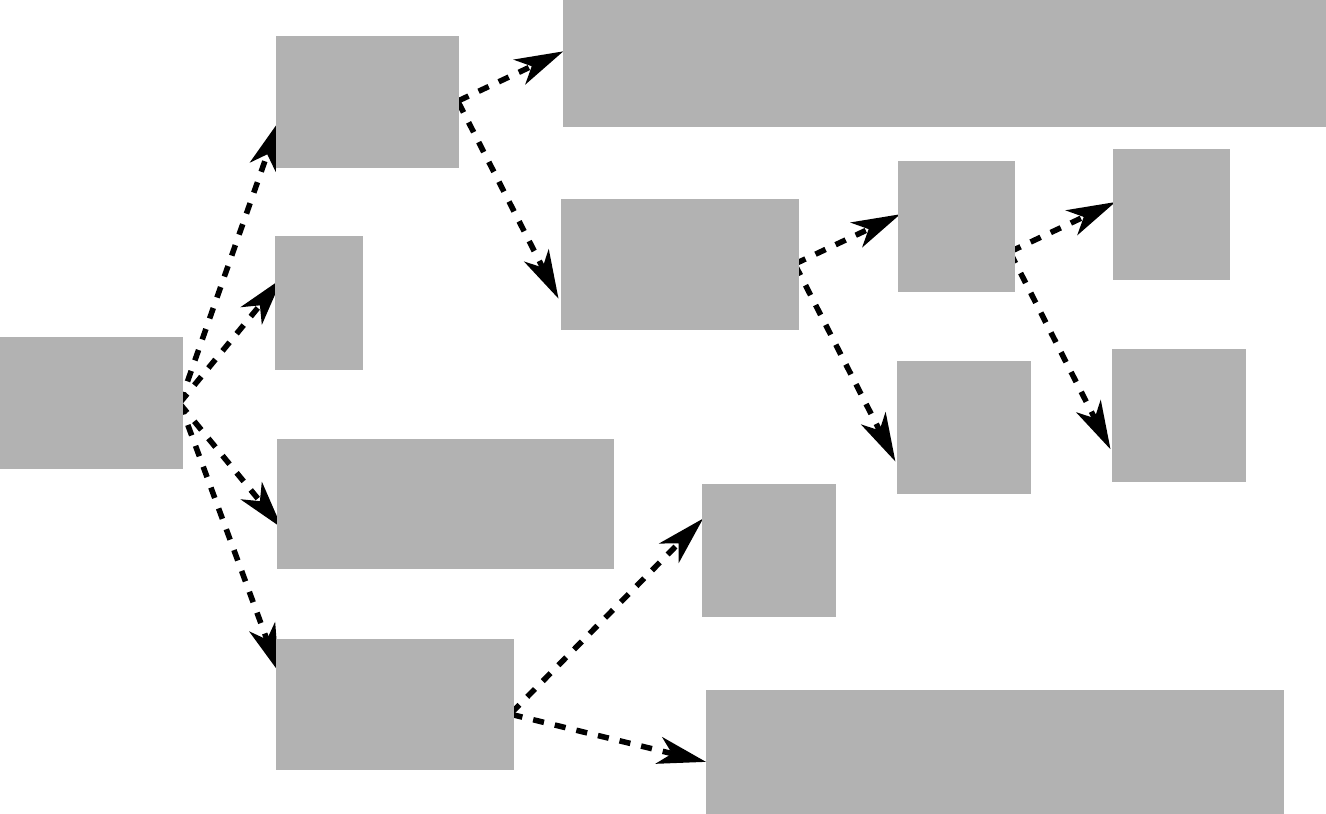}\label{fig:motivation:montecarlo}}
    \hspace{0.03\textwidth}
    \subfigure[Recurrent Neural Network (RNN)]{\includegraphics[width=0.3\textwidth]{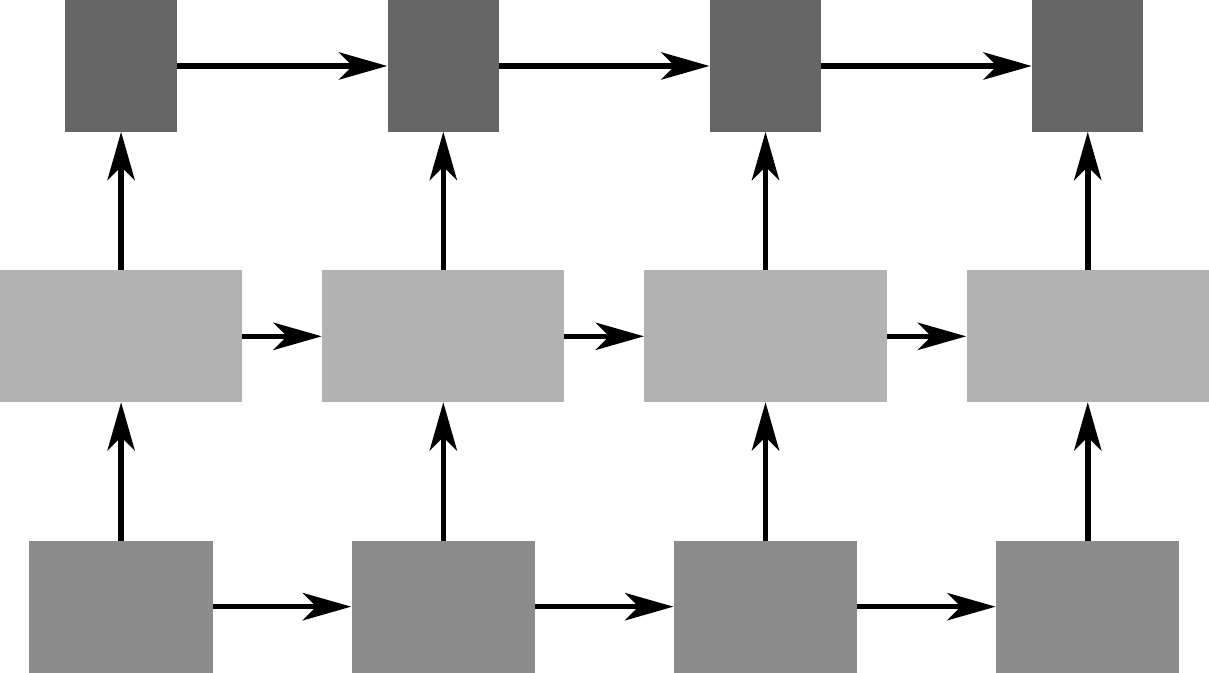}\label{fig:motivation:rnn}}
\end{center}
\vspace{-0.1in}
\caption{
\small{
Example components of a real-time ML application: (a) online processing of streaming sensory
    data to model the environment, (b) dynamic graph construction for Monte
    Carlo tree search (here tasks are simulations exploring sequences of actions), and
    (c) heterogeneous tasks in recurrent neural networks. Different shades represent different types of tasks, and the task lengths represent their durations.
}
}
\label{fig:benefit:grace}
\vspace{-0.1in}
\end{figure*}

Such emerging applications require new levels of programming flexibility and
performance. Meeting these requirements without
losing the benefits of modern distributed execution frameworks (e.g.,
application-level fault tolerance) poses a significant challenge. Our
own experience implementing ML and RL applications in Spark, MPI, and TensorFlow
highlights some of these challenges and gives rise to three groups of requirements
for supporting these applications.
\textit{Though these requirements are critical
for ML and RL applications, we believe they are broadly useful.}

\vspace{0.1cm}
\noindent {\bf Performance Requirements.} Emerging ML applications have stringent
latency and throughput requirements.
\begin{itemize}
\item \textbf{R1:} \emph{Low latency}. The real-time, reactive, and interactive nature of emerging ML
applications calls for fine-granularity task execution with
millisecond end-to-end latency \cite{crankshaw2014missing}.
\item \textbf{R2:} \emph{High throughput}. The volume of micro-simulations
required both for training \cite{gorila} as well as for inference
during deployment \cite{silver2016mastering} necessitates support for high-throughput task execution on
the order of millions of tasks per second.
\end{itemize}

\noindent {\bf Execution Model Requirements.}
Though many existing
parallel execution systems~\cite{mapreduce,spark-cacm16}
have gotten great mileage out of
identifying and optimizing for common computational patterns,
emerging ML applications require far greater flexibility \cite{duan2016benchmarking}.
\begin{itemize}
\item \textbf{R3:} \emph{Dynamic task creation}. RL primitives such as Monte Carlo
tree search may generate new tasks during execution based on the results or the
durations of other tasks.
\item \textbf{R4:} \emph{Heterogeneous tasks}. Deep learning primitives and RL
simulations produce tasks with widely different execution times and resource
requirements.
Explicit system support for heterogeneity of tasks and resources is essential for RL applications.
\item \textbf{R5:} \emph{Arbitrary dataflow dependencies}.
Similarly, deep learning primitives and RL simulations produce arbitrary and
often fine-grained task dependencies (not restricted to bulk synchronous
parallel).
\end{itemize}

\noindent {\bf Practical Requirements.}
\begin{itemize}
  \item  \textbf{R6:} \emph{Transparent fault tolerance}.
  Fault tolerance remains a key requirement for many deployment scenarios, and
  supporting it alongside high-throughput and non-deterministic tasks poses a
  challenge.
  \item  \textbf{R7:} \emph{Debuggability and Profiling}.
  Debugging and performance profiling are the most time-consuming aspects of
  writing any distributed application. This is especially true for ML and RL
  applications, which are often compute-intensive and stochastic.
\end{itemize}

Existing frameworks fall short of achieving one or more of these
requirements~(\secref{sec:relwork}).
We propose a flexible distributed programming model~(\secref{sec:api}) to enable
\textbf{R3-R5}. In addition, we propose a system architecture to support this
programming model and meet our performance requirements~(\textbf{R1-R2})
without giving up key practical requirements~(\textbf{R6-R7}).
The proposed system architecture~(\secref{sec:arch}) builds on two principal components:
a logically-centralized control plane and a hybrid scheduler.
The former enables stateless distributed components and lineage replay.
The latter allocates resources in a bottom-up fashion, splitting locally-born work between
node-level and cluster-level schedulers.

The result is millisecond-level performance on microbenchmarks and
a 63x end-to-end speedup on a representative RL application over a bulk
synchronous parallel (BSP) implementation.

\thispagestyle{empty}
\vspace{-0.25cm}
\section{Motivating Example}
\label{sec:motivation}

\vspace{-0.1cm}

To motivate requirements \textbf{R1-R7}, consider a hypothetical application
in which a physical robot attempts to achieve a goal in an unfamiliar
real-world environment. Various sensors may fuse video and
LIDAR input to build multiple candidate models of the robot's
environment
(Fig.~2a).
The robot is then controlled in
real time using actions informed by a recurrent neural network (RNN) \textit{policy}
(Fig.~2c),
as well as by Monte Carlo tree search (MCTS) and other online
planning algorithms
(Fig.~2b).
Using a physics
simulator along with the most recent environment models, MCTS tries millions of
action sequences in parallel, adaptively exploring the most promising ones.

{\bf The Application Requirements.}
Enabling these kinds of applications involves simultaneously solving a number
of challenges. In this example, the latency requirements~(\textbf{R1}) are
stringent, as the robot must be controlled in real time. High task
throughput~(\textbf{R2}) is needed to support the online simulations for MCTS
as well as the streaming sensory input.

Task heterogeneity~(\textbf{R4}) is present on many scales: some
tasks run physics simulators, others process diverse data streams, and some
compute actions using RNN-based policies. Even similar tasks may exhibit
substantial variability in duration. For example, the RNN consists
of different functions for each ``layer'', each of which may require
different amounts of computation. Or, in a task simulating the robot's actions,
the simulation length may depend on whether the robot
achieves its goal or not.

In addition to the heterogeneity of tasks, the dependencies between tasks
can be complex~(\textbf{R5}, Figs.~2a and~2c)
and difficult to express as batched BSP stages.

Dynamic construction of tasks and their dependencies~(\textbf{R3}) is critical.
Simulations will
adaptively use the most recent environment models as they become
available, and MCTS may choose to launch more tasks exploring
particular subtrees, depending on how promising they are or how fast the
computation is. Thus, the dataflow graph must be constructed
dynamically in order to allow the algorithm to adapt to real-time constraints and
opportunities.

\thispagestyle{empty}
\vspace{-0.25cm}
\section{Proposed Solution}
\label{sec:proposal}

\vspace{-0.1cm}

In this section, we outline a proposal for a distributed execution framework
and a programming model satisfying requirements \textbf{R1-R7} for real-time ML applications.

\vspace{-0.2cm}

\subsection{API and Execution Model}
\label{sec:api}

In order to support the execution model requirements~(\textbf{R3-R5}), we outline
an API that allows arbitrary functions to be specified as remotely executable
tasks, with dataflow dependencies between them.

\begin{enumerate}

    \vspace{-0.2cm}
    \item Task creation is non-blocking. When a \textit{task} is created, a
        \textit{future}~\cite{Baker:futures} representing the eventual return
        value of the task is returned immediately, and the task is executed
        asynchronously.

    \vspace{-0.2cm}
    \item Arbitrary function invocation can be designated as a remote task,
        making it possible to support arbitrary execution kernels~(\textbf{R4}).
        Task arguments can be either
        regular values or futures.  When an argument is a
        future, the newly created task becomes dependent on the task
        that produces that future, enabling arbitrary DAG
        dependencies~(\textbf{R5}).

    \vspace{-0.2cm}
    \item Any task execution can create new tasks without blocking on their
        completion. Task throughput is therefore not limited by the bandwidth
        of any one worker~(\textbf{R2}), and the computation graph is
        dynamically built~(\textbf{R3}).

    \vspace{-0.2cm}
    \item The actual return value of a task can be obtained by calling the
        \texttt{get} method on the corresponding future. This blocks until the
        task finishes executing.

    \vspace{-0.2cm}
    \item The \texttt{wait} method takes a list of futures, a timeout, and a
        number of values. It returns the subset of futures whose
        tasks have completed when the timeout occurs or the requested
        number have completed.

\end{enumerate}

The \texttt{wait} primitive allows developers to specify latency
requirements~(\textbf{R1}) with a timeout, accounting for arbitrarily sized tasks~(\textbf{R4}).
This is important for ML applications, in which a straggler
task may produce negligible algorithmic improvement but block the entire
computation. This primitive enhances our ability to dynamically modify the computation graph
as a function of execution-time properties~(\textbf{R3}).

To complement the fine-grained programming model, we propose using a dataflow execution model
in which tasks become available for execution if and only if their dependencies
have finished executing.

\thispagestyle{empty}
\vspace{-0.3cm}

\subsection{Proposed Architecture}
\label{sec:arch}
\begin{figure}
    \centering
    \includegraphics[width=\columnwidth]{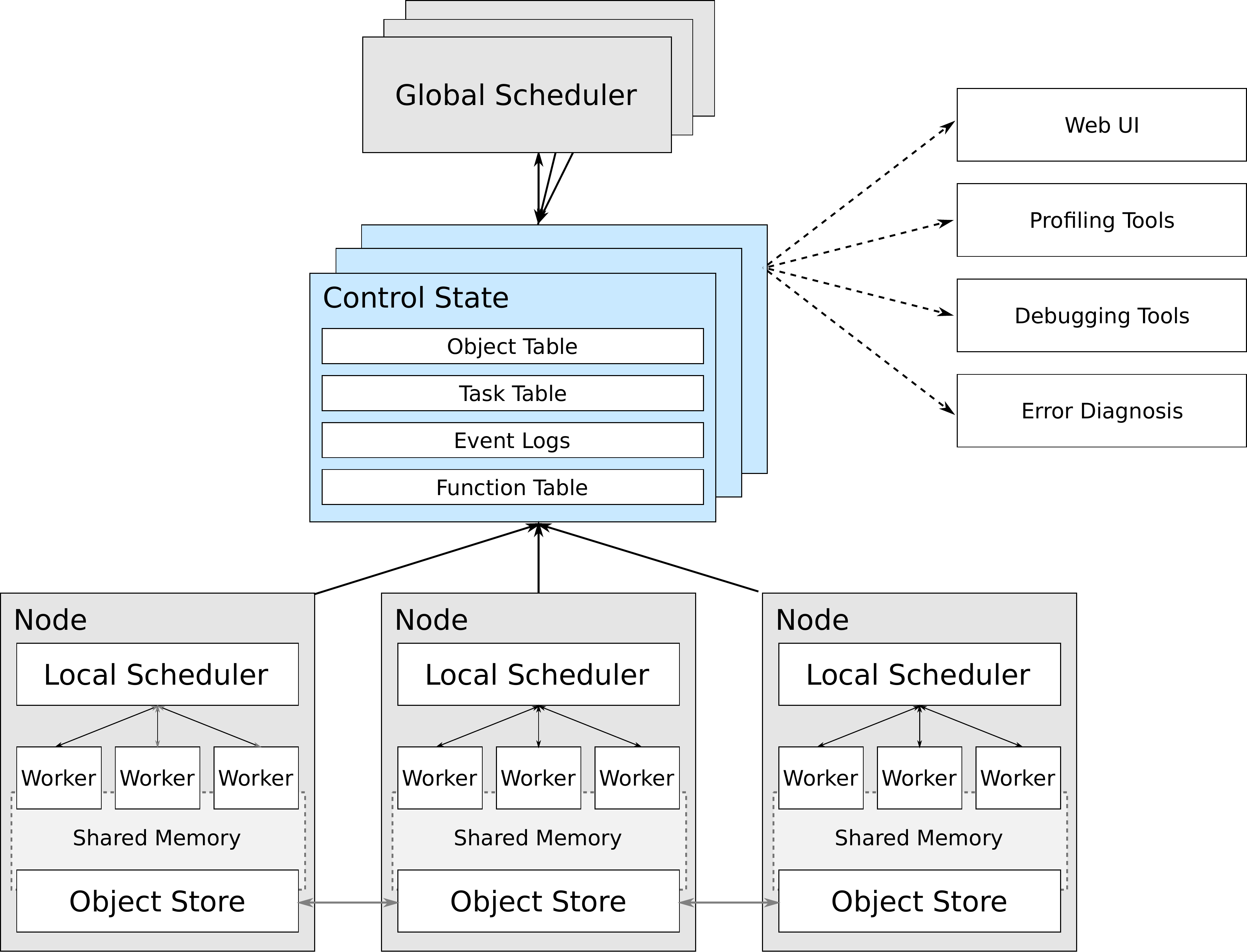}
    \caption{
    	{\small
    	Proposed Architecture, with hybrid scheduling (\secref{sec:arch:hybrid-scheduling}) and a centralized control plane (\secref{sec:arch:control-state}).
        }
    }
    \label{fig:ray-architecture}
\end{figure}

\vspace{-0.2cm}

Our proposed architecture consists of multiple \textit{worker} processes running on each node in the cluster,
one {\em local scheduler} per node, one or more {\em global schedulers}
throughout the cluster, and an in-memory {\em object store} for sharing data between workers (see \figref{fig:ray-architecture}).

The two principal architectural features that enable \textbf{R1-R7}
are a \textit{hybrid scheduler} and a \textit{centralized control
plane}.

\vspace{-0.2cm}

\subsubsection{Centralized Control State}
\label{sec:arch:control-state}

\vspace{-0.2cm}

As shown in \figref{fig:ray-architecture}, our architecture relies on a
logically-centralized control plane~\cite{kreutz2015software}. To
realize this architecture, we use a database that provides both (1) storage for
the system's control state, and (2) publish-subscribe functionality to enable various
system components to communicate with each other.\footnote{In our implementation
we employ Redis \cite{redis2009}, although many other fault-tolerant key-value stores
could be used.}

This design enables virtually any component of the system, except for the
database, to be stateless. So as long as
the database is fault-tolerant, we can recover from component failures by simply
restarting the failed components. Furthermore, the database stores the computation
lineage, which allows us to reconstruct lost data by replaying the
computation~\cite{spark-cacm16}. As a result, this design is fault tolerant
(\textbf{R6}). The database also makes it easy to write tools to profile and inspect the
state of the system (\textbf{R7}).

To achieve the throughput requirement (\textbf{R2}), we shard the database.
Since we require only exact matching operations and since the keys are computed as
hashes, sharding is straightforward. Our early experiments show that
this design enables sub-millisecond scheduling latencies (\textbf{R1}).

\vspace{-0.2cm}

\subsubsection{Hybrid Scheduling}
\label{sec:arch:hybrid-scheduling}

\vspace{-0.2cm}

Our requirements for latency~(\textbf{R1}), throughput~(\textbf{R2}), and
dynamic graph construction~(\textbf{R3}) naturally motivate a hybrid scheduler
in which local schedulers assign tasks to workers or delegate responsibility to
one or more global schedulers.

Workers submit tasks to their local schedulers
which decide to either assign the tasks to other workers on the same physical node or to ``spill over'' the tasks to a
global scheduler. Global schedulers can then assign tasks to
local schedulers based on global information about factors including object
locality and resource availability.

Since tasks may create other tasks, schedulable work may come from any worker in the cluster.
Enabling any local scheduler to handle locally generated work without involving a global scheduler improves
low latency~(\textbf{R1}), by avoiding communication overheads,
and throughput~(\textbf{R2}), by significantly reducing the global scheduler load.
This hybrid scheduling scheme fits well with the recent
trend toward large multicore servers~\cite{fos-wentzlaff}.

\thispagestyle{empty}
\vspace{-0.7cm}

\section{Feasibility}
\label{sec:eval}

\vspace{-0.2cm}

To demonstrate that these API and architectural proposals could in principle
support requirements \textbf{R1-R7}, we provide some simple examples using the
preliminary system design outlined in \secref{sec:proposal}.

\subsection{Latency Microbenchmarks}

Using our prototype system, a task can be created, meaning that
the task is submitted asynchronously for execution and a future is returned, in
around $35\mu$s. Once a task has finished executing, its return value can be
retrieved in around $110\mu$s. The end-to-end time, from
submitting an empty task for execution to retrieving its return value,
is around $290\mu$s when the task is scheduled locally and
$1$ms when the task is scheduled on a remote node.

\subsection{Reinforcement Learning}

We implement a simple workload in which an RL agent is trained to play an Atari
game. The workload alternates between stages in which actions are taken in
parallel simulations and actions are computed in parallel on GPUs. Despite the
BSP nature of the example, an implementation in Spark is \textbf{9x} slower
than the single-threaded implementation due to system overhead. An
implementation in our prototype is \textbf{7x} faster than the single-threaded version
and \textbf{63x} faster than the Spark implementation.\footnote{In this comparison,
the GPU model fitting could not be naturally parallelized on Spark, so the numbers are
reported as if it had been perfectly parallelized with no overhead in Spark.}

This example exhibits two key features. First, tasks are very small (around
7ms each), making low task overhead critical.  Second, the tasks are
heterogeneous in duration and in resource requirements (e.g., CPUs and
GPUs).

This example is just one component of an RL workload, and would typically
be used as a subroutine of a more sophisticated (non-BSP) workload. For
example, using the {\tt wait} primitive, we can
adapt the example to process the simulation tasks in the order that they finish
so as to better pipeline the simulation execution with the action computations
on the GPU, or run the entire workload nested within a larger adaptive
hyperparameter search. These changes are all straightforward using the API
described in \secref{sec:api} and involve a few extra lines of code.

\thispagestyle{empty}
\section{Related Work}
\label{sec:relwork}

\textbf{Static dataflow
systems}~\cite{mapreduce,spark-cacm16,dryad,murray:naiad} are well-established
in analytics and ML, but they require the dataflow graph
to be specified upfront, e.g., by a driver program. Some, like
MapReduce~\cite{mapreduce} and Spark~\cite{spark-cacm16}, emphasize BSP
execution, while others, like Dryad~\cite{dryad} and Naiad~\cite{murray:naiad},
support complex dependency structures~(\textbf{R5}). Others, such as
TensorFlow~\cite{tensorflow-osdi16} and MXNet~\cite{mxnet-learningsys},
are optimized for deep-learning workloads. However, none of
these systems fully support the ability to
dynamically extend the dataflow graph in response to both input data and task
progress~(\textbf{R3}).

\textbf{Dynamic dataflow systems} like CIEL~\cite{murray:ciel} and Dask~\cite{dask-scipy15} support many
of the same features as static dataflow systems, with additional support for
dynamic task creation~\textbf{(R3)}. These systems meet our execution model
requirements~(\textbf{R3-R5}). However, their architectural limitations, such
as entirely centralized scheduling, are such that low latency~(\textbf{R1})
must often be traded off with high throughput~(\textbf{R2}) (e.g., via
batching), whereas our applications require both.

\textbf{Other systems} like Open~MPI~\cite{openmpi} and actor-model variants
Orleans~\cite{bykov2011orleans} and Erlang~\cite{armstrong1993concurrent} provide
low-latency~(\textbf{R1}) and high-throughput~(\textbf{R2}) distributed
computation.
Though these systems do in principle provide primitives for supporting our execution
model requirements~(\textbf{R3-R5})
and have been used for ML \cite{coates2013deep,amodei2015deep},
much of the logic required for
systems-level features, such as fault tolerance~(\textbf{R6}) and
locality-aware task scheduling, must be implemented at the application level.

\thispagestyle{empty}
\section{Conclusion}
\label{sec:conclusion}

Machine learning applications are evolving to require dynamic dataflow parallelism
with millisecond latency and high throughput, posing a severe challenge for existing frameworks.
We outline the requirements for supporting this emerging class of real-time ML applications,
and we propose a programming model and architectural design to address
the key requirements~(\textbf{R1-R5}), without compromising existing requirements~(\textbf{R6-R7}).
Preliminary, proof-of-concept results confirm millisecond-level system overheads
and meaningful speedups for a representative RL application.

\thispagestyle{empty}
\subsubsection*{Acknowledgments}

We would like to thank Richard Shin for substantial contributions to the
development of our prototype.

This research is supported in part by DHS Award HSHQDC-16-3-00083, NSF CISE
Expeditions Award CCF-1139158, and gifts from Ant Financial, Amazon Web
Services, CapitalOne, Ericsson, GE, Google, Huawei, Intel, IBM, Microsoft and
VMware.

\pagebreak
\thispagestyle{empty}
{\footnotesize \bibliographystyle{acm}
\bibliography{paper}}


\end{document}